\documentclass[fleqn,12pt,twoside]{article}
\usepackage{espcrc1}

\usepackage{graphicx}

\newcommand{\R}{{\cal R}}

\title{The universal freeze-out criterion at SPS and RHIC}

\author{Boris Tom\'a\v sik and Urs Achim Wiedemann \\[2ex]
        CERN, Theory Division, CH-1211 Geneva 23, Switzerland}

\begin{document}

\maketitle

\begin{abstract}
We formulate a freeze-out criterion for ultra-relativistic heavy
ion collisions in terms of the pion escape probability from the 
collision region. We find that the increase in pion phase-space 
density from SPS to RHIC reported at this conference has a 
small influence on particle freeze-out because of 
the small $\pi\pi$ cross-section. Our treatment takes into account
dynamical expansion, chemical composition
and momentum of the escaping particle for particle freeze-out.
It supports a freeze-out at rather low temperature---below 100~MeV---and 
earlier decoupling of high~$p_\perp$ particles.
\end{abstract}
\\

A precise formulation of the condition under which particles
decouple (``freeze-out'') from a heavy ion collision is important 
for understanding the dynamical origin of the final state. The
observation of STAR \cite{STARhere} that the
pion phase-space density increases at RHIC indicates that 
freeze-out is not characterised solely by a universal 
value of {\em pion phase-space density}, in contrast to a
recent suggestion \cite{Ferenc:1999ku}. Also, freeze-out is 
not characterised solely by a universal value of the 
{\em total particle density} since the volume
from which particles decouple  shows a non-monotonous
dependence on collision energy \cite{CERES}. 

It has been argued \cite{Nagamiya:tf} and
demonstrated \cite{CERES} that the chemical composition 
of the system must be taken into account. 
For pion freeze-out, the nucleon and anti-nucleon density 
is more important than that of pions since the pion-nucleon
cross-section is larger than the pion-pion one. With increasing 
collision energy the pion 
abundance and their contribution to scattering grows, 
but it is argued that the {\em momentum-averaged}
mean free path of a pion at freeze-out stays constant \cite{CERES}.

This argumentation still ignores the influence of global dynamics on
the freeze-out criterion. A strong expansion makes the particle density 
decrease fast and if the characteristic time scale for density
decrease $\tau_{\rm exp} = (\partial_\mu u^\mu)^{-1}$ is smaller 
than the mean free time between collisions $\tau_{\rm scatt}$ a
particle is likely to seize interacting, i.e., to freeze-out 
\cite{Bondorf:kz}.

In our work we study---in addition to these two aspects---how the
probability for a particle to decouple depends on its momentum
\cite{Tomasik:2002qt}. We formulate freeze-out as condition for 
an individual particle: a {\em particle freezes-out} at the time
of its last scattering. 
This is in contrast to the ``fireball freeze-out'' which
is understood as a boundary in space-time between interacting matter
and free-streaming particles. In the extreme case of a hydrodynamic 
model the ``fireball freeze-out'' is usually treated by the 
Cooper-Frye prescription \cite{Cooper:1974mv} in which {\em all}
particles decouple along the same three-dimensional hyper-surface
when {\em the medium} fulfils certain condition. 
Our treatment goes beyond Cooper-Frye: particle freeze-out
depends on its {\em type, position, and momentum} in addition to the 
characteristics of the medium. This corresponds to  
freeze-out as described by a cascade event generator.

The freeze-out criterion which we explore is based on the {\em escape 
probability} \cite{Sinyukov:2002if} 
\begin{equation}
\label{escprob}
     {\cal P}(x,p,\tau) = 
     \exp \left ( - \int_\tau^\infty \, d\bar{\tau} \, 
         \R(x+v\bar{\tau},p) \right )\, ,
\end{equation}
which determines the probability that a pion emitted with momentum
$p$ from the position $(x,\, \tau)$ escapes from the medium without
further interaction. Here, the scattering rate $\R(x,p)$ is integrated 
along the trajectory of the pion. Freeze-out is assumed to occur when
${\cal P}(x,p,\tau)$ reaches a {\em universal} value of order one. 

\underline{The role of collective expansion:}
To illustrate how collective expansion affects particle freeze-out,
we consider the case of particles with vanishing momentum in the
centre of a fireball with longitudinally boost-invariant 
and transversely  linear expansion velocity profile of the form
\begin{eqnarray}
u^\mu & = & (\cosh\eta\, \cosh\eta_t,\, \cos\phi\, \sinh\eta_t,\,
\sin\phi\, \sinh\eta_t,\, \sinh\eta\, \cosh\eta_t)\, .
\end{eqnarray} 
Here, $\eta = {\rm Arctanh}(z/t)$ and $\eta_t  = \xi r$ where 
$\xi=0.08\,\mbox{fm}^{-1}$ is a typical value for the 
transverse expansion gradient \cite{Tomasik:1999cq}. At the time
$\tau_{\rm em}$ at which the pion is emitted, the density of 
scattering centres decreases by the rate
\begin{equation}
\label{denrate}
\left . -\frac{1}{\rho} \frac{\partial\rho}{\partial \tau}
\right |_{\tau = \tau_{\rm em}} =
\partial_\mu u^\mu = \frac{1}{\tau_{\rm em}} + 2\xi \, .
\end{equation}
A power-law decrease of density with time is
consistent with (\ref{denrate}) if
\begin{equation}
\label{power}
\rho(\tau) = \rho_{\rm em}\left ( \frac{\tau_{\rm em}}{\tau} \right )^\alpha
\, ,
\qquad \alpha = 1+2\xi\tau_{\rm em}\, .
\end{equation}
Under the assumption $\R\propto \rho$ we obtain for the opacity integral
of a particle of zero momentum
\begin{equation}
\int_{\tau_{\rm em}}^\infty d\tau\, \R(\tau) 
= \frac{\R_{\rm em}}{\alpha -1} \tau_{\rm em} = \frac{\R_{\rm em}}{2\xi}\, ,
\label{rflow}
\end{equation}
where $\R_{\rm em}$ is the scattering rate at $\tau_{\rm em}$.
Technically, the calculation of this opacity integral for particles
of non-zero momentum $p$ is more complicated, since one has to follow
the propagation of the particle through layers of different density.
For a quantitative statement, this requires a model of the space-time
evolution of the fireball. Qualitatively, however, the general result
will be consistent with the main feature of the case $p=0$ considered 
here: as seen from (\ref{rflow}), freeze-out from a denser fireball 
(larger $\R_{\rm em}$) is possible if it is compensated by a stronger 
expansion gradient such that the value of opacity integral stays
constant.

\underline{The role of chemical composition:}
To determine the value of $\R_{\rm em}$ for
collisions at $\sqrt{s}=17\, A\mbox{GeV}$ (SPS) and 
$\sqrt{s} = 130\, A\mbox{GeV}$ (RHIC), we calculate 
the scattering rate 
\begin{equation}
\R(p) =  \sum_i \int d^3k \, \rho_i(k) \, \sigma_i^\prime(s) \, 
|v_\pi- v_i |\, ,
\label{scrate}
\end{equation}
where $\sigma_i^\prime$ is the cross-section for pion scattering
on species $i$. For the collinear cross-section we use
the parametrisation
\begin{equation}
\label{cross}
\sigma_i(\sqrt{s})  =  \sum_{r} 
      \langle j_i, m_i, j_\pi, m_\pi ||  J_r, M_r \rangle 
      \frac{2S_r + 1}{(2S_i + 1)(2S_\pi +1)} 
       \frac{\pi}{p^2_{\rm CMS}}
         \frac{\Gamma_{r\to \pi i}\Gamma_{i,{\rm tot}}}{%
           (M_r - \sqrt{s})^2 + \Gamma^2_{i,{\rm tot}}/4} \, ,
\end{equation}
where the sum runs over all resonances in the $\pi i$ scattering.
Particle densities $\rho_i(k)$ are assumed to be thermal. For
realistic temperatures, only low-lying resonances are relevant and we 
include $i=\pi,\, N,\, \bar N,\, K,\, \rho,\, \Delta,\, \bar\Delta$
when summing in (\ref{scrate}) over scattering partners. Since the 
freeze-out temperature is not known a priori, we scan three values:
$T=90,\, 100,\, 120\, \mbox{MeV}$. Chemical potentials for 
pions are estimated from data on phase-space density, those for other 
species from measured ratios of $dN/dy$ at mid-rapidity (see 
\cite{Tomasik:2002qt} for more details on the estimates). 

%
\begin{figure}[t]
\begin{center}
\includegraphics[scale=1.09]{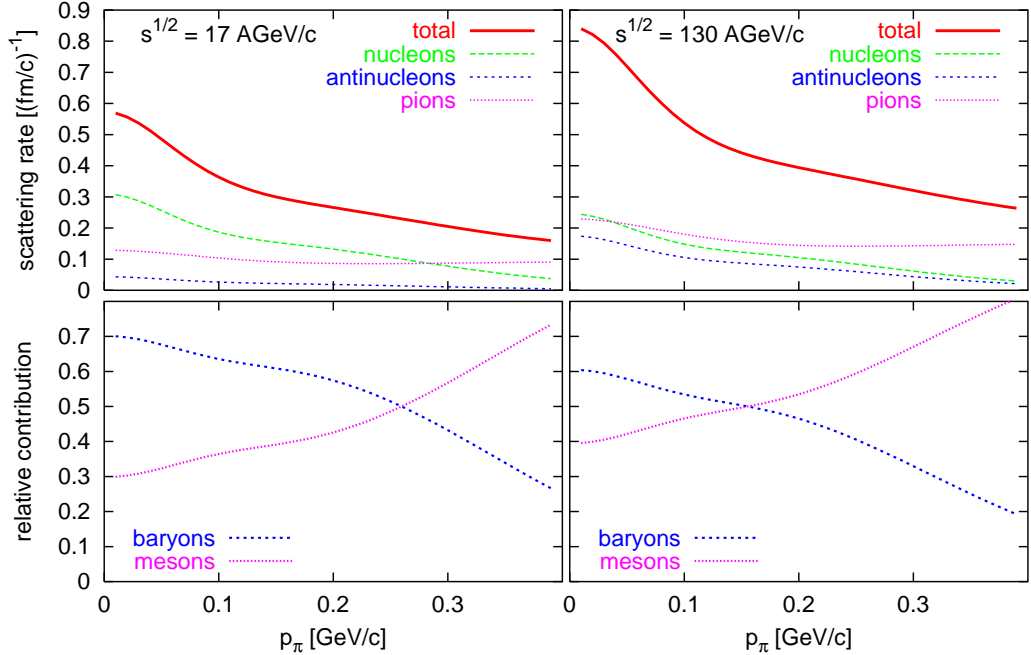}
\vspace{-1cm}
\caption{The pion scattering rate as a function of pion momentum
with respect to medium
calculated at $T=100\, \mbox{MeV}$ and 
the highest estimates of chemical potentials allowed by data
from SPS (left column)
and RHIC (right column). Contributions to the total scattering rate 
from scattering on nucleons, anti-nucleons and pions are indicated.
The lower row shows the baryonic and mesonic relative contributions. 
\label{F:contribs}\vspace{-0.9cm}
}
\end{center}
\end{figure}
%
Figure~\ref{F:contribs} shows the scattering rates as a function
of the pion momentum {\em relative to the heat bath} calculated 
for SPS and RHIC, as well as the most important contributions to 
$\R$. In spite of the strong increase of the pion phase-space density 
at RHIC \cite{STARhere}, the relative meson 
contribution to the total scattering rate does not grow proportionally
since the $\pi\pi$ cross-section is much smaller than the one for
$\pi N$. The total baryonic contribution changes very little since
the smaller amount of baryons at RHIC is roughly compensated by 
anti-baryons. In summary, there is about 10\%
of the relative contribution shifted from baryons to mesons when going
from SPS to RHIC.

%
\begin{figure}[t]
\begin{center}
\includegraphics[scale=1.3]{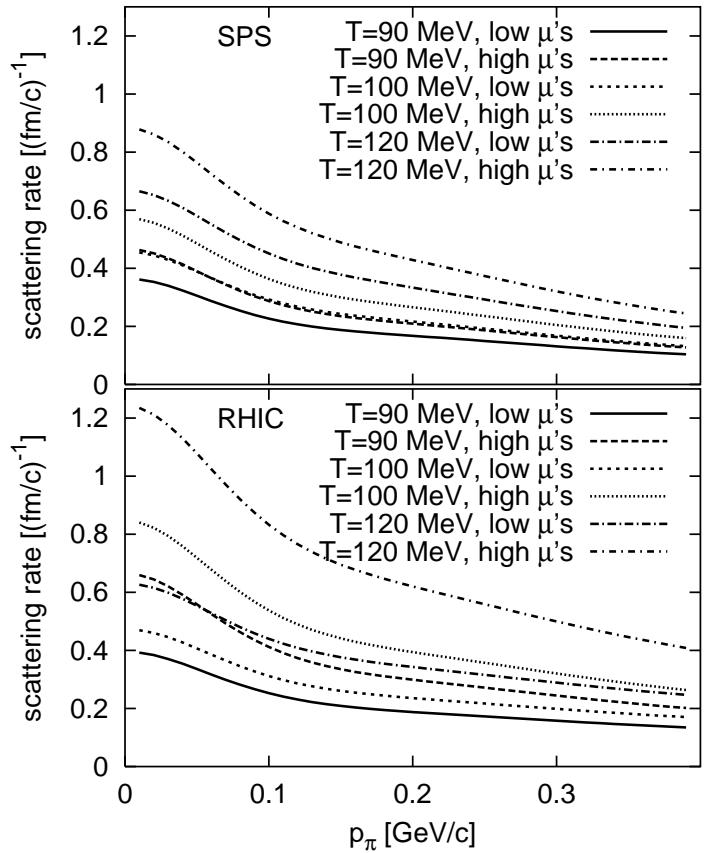}
\hspace{1em}
\begin{minipage}[b]{5.4cm}
\caption{The pion scattering rate as a function of pion momentum
with respect to the medium
calculated for different temperatures and a range of chemical potentials
allowed by data (see \protect\cite{Tomasik:2002qt}). New results of
STAR collaboration \protect\cite{STARhere} correspond rather
to the ``high $\mu$'s'' estimates.
\label{F:spsrhic}}
\end{minipage}
\end{center}
\end{figure}
%
As seen in Figure~\ref{F:spsrhic}, the scattering rate depends strongly 
on the assumed freeze-out temperature since the pion-nucleon scattering 
is typically dominated by total CMS energy below the $\Delta$
resonance peak. At higher temperature, average CMS energies
move closer to the peak value thus leading to increased scattering
cross-sections. Estimating the pion escape probability (\ref{escprob})
from eq.~(\ref{rflow}) and $\xi = 0.08\, \mbox{fm}^{-1}$, one finds
that a pion only has a reasonable chance to escape (of order 50\%)
when the temperature is 100~MeV at most.

We finally point out that the scattering rates in Figure~\ref{F:spsrhic}
generically decreases with increasing pion momentum. This leads us to 
conjecture that high $p_\perp$ pions may decouple earlier, from hotter 
and smaller system. This may provide a novel as yet unexplored
contribution to the observed $M_\perp$ dependence of HBT radii.

The freeze-out criterion studied here differs from the Cooper-Frye
formalism typically employed in hydrodynamical model studies. 
It will be interesting to explore to what extent this affects
the predictions of hydrodynamical simulations for the collision
evolution.

\end{document}